\documentclass[aps,twocolumn,prb,amsmath,amssymb,superscriptaddress]{revtex4-1}

\usepackage{amsmath}
\usepackage{amssymb}
\usepackage[varg]{txfonts}
\usepackage{color}
\usepackage[colorlinks]{hyperref}
\usepackage{tikz}
\usetikzlibrary{shapes}
\usepackage{xcolor}

\usepackage{graphicx}
\usepackage{bm}

\begin{document}
\date{\today}
\title{Engineering nonlinear spin-orbit torque driven by intra-band transport in MoSe$_2$/CrI$_3$ and WSe$_2$/CrI$_3$ van der Waals heterostructures}
\author{Leyla Majidi}
\email{l.majidi@scu.ac.ir}
\affiliation{Faculty of Science, Department of Physics, Shahid Chamran University of Ahvaz, Ahvaz, Iran}
\author{Azadeh Faridi}
\affiliation{School of Quantum Physics and Matter, Institute for Research in Fundamental Sciences (IPM), Tehran 19395-5531, Iran}
\author{Reza Asgari}
\affiliation{Department of Physics, Zhejiang Normal University, Jinhua, Zhejiang 321004, China}
\affiliation{School of Quantum Physics and Matter, Institute for Research in Fundamental Sciences (IPM), Tehran 19395-5531, Iran}
\date{\today}

\begin{abstract}
The role of nonlinear carrier dynamics in current-driven spin phenomena remains poorly understood in van der Waals magnetic heterostructures. Here, we investigate the spin polarization and the resulting spin-orbit torque (SOT) in transition-metal dichalcogenide/chromium iodide (TMDC/CrI$_3$) heterostructures, focusing on WSe$_2$/CrI$_3$ and MoSe$_2$/CrI$_3$, in the zero-Rashba spin-orbit coupling (SOC) limit and beyond the linear-response regime. We find that linear spin polarization is forbidden by symmetry, making nonlinear intraband transitions the dominant mechanism in this regime. Consequently, the current-driven spin response differs fundamentally from conventional linear-response predictions. The resulting nonlinear spin polarization generates a purely field-like torque whose magnitude and sign depend sensitively on the chemical potential and doping type, with pronounced asymmetries between n- and p-doped systems. A strong material dependence is also observed: n-doped MoSe$_2$/CrI$_3$ exhibits a torque nearly an order of magnitude larger than that of WSe$_2$/CrI$_3$, together with two sign reversals, reflecting differences in their SOC and proximity-exchange parameters. Structural and electrostatic control through the twist angle and gate electric field provides additional means of tuning the nonlinear torque, including substantial modulation and controllable sign reversals. These findings establish TMDC/CrI$_3$ bilayers as a versatile platform for engineering nonlinear spin-orbit phenomena in ultrathin, low-power spintronic devices.
\end{abstract}

\maketitle
\section{introduction}

Spin-orbit torque (SOT) is a cutting-edge mechanism in spintronics that enables efficient electrical control of magnetization. Compared with spin-transfer torque (STT)~\cite{slonczewski1996current,berger1996emission,myers1999current,katine2000current,Zare2017,Majidi2018}, which arises from the transfer of spin angular momentum from a spin-polarized current to a magnetic layer, SOT provides distinct advantages, including high-speed operation, low power consumption, and enhanced device reliability, making it a promising approach for next-generation magnetic random-access memory (MRAM). A typical SOT device consists of a material with strong spin-orbit coupling (SOC) coupled to a magnetic layer. When an electric current flows through the SOC-active material, spin-dependent interactions generate a nonequilibrium spin density, which exerts a torque on the magnetization and thereby enables electrical manipulation of the magnetic order~\cite{gambardella2011current,manchon2019current}.

Recent studies have explored a wide range of materials to enhance SOT efficiency. Heavy metals (HMs), such as Pt, W, and Ta, have been extensively employed because of their strong SOC and associated spin-charge conversion mechanisms~\cite{liu2012spin,liu2012current,pai2014enhancement,skinner2015complementary}. More recently, two important developments have significantly broadened the materials landscape for SOT devices. First, topological insulators (TIs), including Bi- and Sb-based chalcogenides, have demonstrated exceptionally large SOT efficiencies, which have been attributed primarily to the spin-momentum locking of their topological surface states~\cite{fan2014magnetization,wang2015topological,kondou2016fermi,han2017room}. Second, transition-metal dichalcogenides (TMDCs) have emerged as promising platforms for generating sizable SOT in TMDC/ferromagnet (FM) bilayers, owing to their strong intrinsic SOC and rich symmetry properties~\cite{liu2020two,hidding2020spin,tang2021spin,galceran2021control}. These characteristics provide new opportunities for developing ultrathin and energy-efficient spintronic devices.

The broad family of TMDCs encompasses materials with diverse crystal symmetries and electronic structures, giving rise to both conventional and unconventional SOT components with distinct orientations. An increasing number of experimental studies have demonstrated SOT in a variety of TMDC-based heterostructures, including WTe$_2$~\cite{macneill2017control,li2018spin,zhao2020observation}, TaTe$_2$~\cite{stiehl2019current,hoque2020charge}, MoS$_2$~\cite{zhang2016research,shao2016strong}, WSe$_2$~\cite{shao2016strong,novakov2021interface,hidding2021interfacial}, NbSe$_2$~\cite{guimaraes2018spin}, and TaS$_2$~\cite{husain2020large}. These systems span semimetallic, semiconducting, and metallic regimes and have been integrated with a variety of ferromagnetic layers. Importantly, these studies demonstrate that the ferromagnetic layer can actively influence the magnitude and symmetry of the SOT, rather than merely serving as a passive recipient of interfacial spin currents~\cite{dolui2020spin}.

A central challenge for the practical implementation of SOT-based devices is the precise and versatile control of both the magnitude and orientation of the torque. Such controllability is essential for optimizing device performance and improving the reliability and functionality of spintronic applications. Although electrically tunable SOTs have been demonstrated in HM/FM~\cite{mishra2019electric}, TI/FM~\cite{han2017room}, and TMDC/FM~\cite{lv2018electric,lin2023magnetization} bilayers, as well as unconventional SOTs induced by strain~\cite{guimaraes2018spin}, achieving robust external control over the torque remains challenging. In this regard, TMDC/CrI$_3$ heterostructures provide a particularly promising platform because the proximity-induced exchange interaction in TMDC monolayers, such as MoSe$_2$ and WSe$_2$, can be strongly modified by electrostatic gating and twist angle~\cite{Zollner19}. This tunability offers an effective route for controlling the electronic and magnetic properties relevant to current-induced spin phenomena in these heterostructures~\cite{majidi2022electrical,Zollner23}.

CrI$_3$ is a prototypical two-dimensional (2D) magnetic material that exhibits ferromagnetic order in the monolayer limit and antiferromagnetic interlayer coupling in bilayer structures. These unusual magnetic properties have stimulated extensive investigations of its fundamental and functional characteristics~\cite{jiang2018controlling,kim2019exploitable,lei2021magnetoelectric,zhang2022all}, as well as its integration with TMDCs    ~\cite{zhong2017van,seyler2018valley,hu2020manipulation,ge2022enhanced,dolui2020proximity,heibenbuttel2021valley,da̧browski2022all}. The strong interfacial proximity effect in these van der Waals (vdW) heterostructures provides an efficient means of transferring and controlling magnetic information between the constituent layers. In particular, numerous experimental and theoretical studies have demonstrated substantial control of valley and spin degrees of freedom in TMDC/CrI$_3$ bilayers, especially in WSe$_2$-based heterostructures~\cite{majidi2022electrical,Zollner23,zhong2017van,seyler2018valley,hu2020manipulation,ge2022enhanced,Majidi141,Majidi142}. In addition, all-optical switching of magnetization in CrI$_3$ thin films~\cite{zhang2022all} and optical control of spin in TMDC/CrI$_3$ heterostructures~\cite{da̧browski2022all} have further highlighted the rich possibilities for manipulating magnetic and spin degrees of freedom in these systems.

More recently, current-driven magnetic control has also been demonstrated in vdW heterostructures containing CrI$_3$. In particular, a proximity-induced SOT was predicted in TaSe$_2$/bilayer-CrI$_3$, where the current-induced torque can align the magnetization of the bottom CrI$_3$ layer with that of the top layer, thereby converting the antiferromagnetic configuration of bilayer CrI$_3$ into a ferromagnetic one~\cite{dolui2020proximity}. This result highlights the potential of TMDC/CrI$_3$ heterostructures for current-driven manipulation of magnetic order and motivates further investigation of the microscopic mechanisms governing SOT in these systems.

In Rashba-active vdW magnetic heterostructures, nonlinear contributions to current-induced spin polarization are typically strongly suppressed compared with the linear response~\cite{Majidi26}, raising the fundamental question of whether nonlinear intraband response becomes the leading contribution. Here, we address this question by developing a microscopic theory of current-induced SOT in TMDC/CrI$_3$ heterostructures, focusing on WSe$_2$/CrI$_3$ and MoSe$_2$/CrI$_3$ in the zero-Rashba SOC limit and beyond the linear-response regime. Our primary objective is to determine how the magnitude, sign, and symmetry of the torque can be controlled through external parameters, including electrostatic gating and interlayer twisting, thereby providing a route toward electrically programmable, ultrathin spintronic devices.

Using a single-band steady-state Boltzmann approach, we show that neither intrinsic interband nor extrinsic intraband transitions contribute to the spin polarization in the linear-response regime. Instead, the spin polarization arises entirely from nonlinear intraband processes associated with electric-field-induced modifications of the carrier distribution. Importantly, this nonlinear contribution is not merely a small higher-order correction; it becomes the leading-order physical mechanism when the linear response is symmetry-forbidden and consequently determines the magnitude, sign, and symmetry of the resulting torque. The nonlinear spin polarization generates a strong in-plane, field-like SOT acting on the magnetization of the CrI$_3$ layer. The torque exhibits a pronounced dependence on the chemical potential, including multiple sign reversals and a marked asymmetry between n- and p-type doping. In particular, the n-doped MoSe$_2$/CrI$_3$ bilayer exhibits a torque up to an order of magnitude larger than that of WSe$_2$/CrI$_3$, accompanied by two sign reversals. This enhanced and nonmonotonic response originates from the negative sign and smaller magnitude of the conduction-band SOC parameter in MoSe$_2$.

We further demonstrate that interlayer twisting provides an effective means of controlling the nonlinear SOT. Depending on the chemical potential, twisting the TMDC relative to CrI$_3$ can either enhance or suppress the torque and can induce or eliminate sign reversals in both WSe$_2$/CrI$_3$ and MoSe$_2$/CrI$_3$. A particularly strong enhancement occurs in p-doped WSe$_2$/CrI$_3$. Electrostatic gating provides an additional degree of control, yielding nearly an order-of-magnitude modulation of the torque in n-doped WSe$_2$/CrI$_3$ and enabling a reversal of its sign in the p-doped regime. More generally, gate fields enable simultaneous control of the sign and magnitude of the SOT over a broad range of twist angles in both p- and n-doped structures. Together, these results establish doping, interlayer twist, and electrostatic gating as powerful and complementary control parameters for engineering nonlinear SOT in TMDC/CrI$_3$ heterostructures.

The structure of the paper is as follows. Section \ref{Model} introduces the theoretical model and the fundamental formalisms used to analyze the current-induced nonlinear spin-orbit torque in a twisted TMDC/CrI$_3$ bilayer. Section \ref{results} presents the numerical results for the induced non-equilibrium spin polarization and the resulting spin-orbit torque acting on the magnetization of the ferromagnetic layer. A brief summary of our main findings is provided in Sec. \ref{conclusion}.

\section{Model and theory}\label{Model}

We consider a wide TMDC/CrI$_3$ bilayer oriented normal to the $z$ axis, with a monolayer CrI$_3$ acting as a 2D ferromagnet for $z<0$ and a monolayer WSe$_2$ or MoSe$_2$ serving as the TMDC for $z>0$. The magnetic insulator CrI$_3$, which exhibits out of plane ferromagnetism in the monolayer, is weakly coupled to the TMDC through vdW interactions, thereby preserving the characteristic electronic band structure of the TMDC. In pristine TMDC monolayers, the K and K' valleys are degenerate as required by time reversal symmetry. Proximity to the 2D magnetic material CrI$_3$  provides a direct and efficient mechanism to lift this valley degeneracy via the magnetic proximity effect, which induces an exchange-driven spin-splitting in the TMDC bands in addition to the intrinsic (valley Zeeman) SOC-induced splitting. In this configuration, the magnetization direction of the TMDC aligns with that of the iodine atoms and opposes that of the chromium atoms, resulting in negative proximity-exchange parameters when the net magnetization of CrI$_3$ points along the positive $z$ direction toward the TMDC. The effective low-energy Hamiltonian describing a TMDC monolayer in the presence of proximity exchange, following Ref.~\onlinecite{Zollner19}, takes the form

\begin{eqnarray}
\label{H}
\mathcal{H}&=&\hbar v_{\rm F}\hat{s}_0\otimes(\tau {\hat{\sigma}}_x k_x+{\hat{\sigma}}_y k_y)+\frac{\Delta}{2}{\hat{s}}_0\otimes{\hat{\sigma}}_{z}+\tau {\hat{s}}_z\otimes(\lambda_c{\hat{\sigma}}_{+}\nonumber\\
&+&\lambda_v{\hat{\sigma}}_{-})+(\hat{\bm{s}}.{\bm{m}})\otimes(B_c{\hat{\sigma}}_{+}+B_v{\hat{\sigma}}_{-}).
\end{eqnarray}

The valley index $\tau=\pm1$ labels the K and K' points, and $v_{\rm F}$ denotes the Fermi velocity. The pseudospin Pauli matrices $\hat{\sigma}_i$ ($i = 0,x,y,z$) act on the conduction  and valence band subspaces, while $\hat{s}_i$ ($i = 0,x,y,z$) operate on the real spin degree of freedom. The parameter $\Delta$  represents the orbital band gap. The spin splittings of the conduction- (valence-) band arising from intrinsic SOC and proximity-induced exchange field are characterized by the parameters $\lambda_{c(v)}$ and $B_{c(v)}$, respectively. The local magnetization of the ferromagnetic CrI$_3$ layer is described by ${\bm{m}}=(m_x,m_y,m_z)=(\sin\theta_m\cos\phi_m,\sin\theta_m\sin\phi_m,\cos\theta_m)$, where $\phi_m $ and $\theta_m$ specify its azimuthal and polar orientations. For compact notation, we introduce ${\hat{\sigma}}_{\pm} = ({\hat{\sigma}}_0 \pm {\hat{\sigma}}_z)/2$. Using the parameters reported in Ref.~\onlinecite{Zollner19}, we set $\Delta= 1.327 (1.301)$ eV, $v_{\rm F}= 5.845\times10^5 (4.591\times10^5)$ m/s, $\lambda_{c}=13.81 (-9.678)$ meV, $\lambda_{v}=240.99 (94.43)$ meV, $B_c= -1.783 (-1.592)$ meV and $B_v= -1.583 (-1.426)$ meV for the WSe$_2$(MoSe$_2$)/CrI$_3$ bilayer.

In the WSe$_2$/CrI$_3$ bilayer, the negative proximity-induced exchange field shifts the spin-up subbands downward and the spin-down subbands upward in both the conduction and valence bands by the amounts $|B_c|$ and $|B_v|$, respectively. As a result, the spin-split subbands in the K valley move closer together, whereas those in the K' valley are pushed farther apart. In contrast, the MoSe$_2$/CrI$_3$ bilayer exhibits a qualitatively different behavior due to the opposite sign of $\lambda_c$. Here, the proximity-induced exchange field brings both spin-up subbands of the two valleys closer together, and likewise brings the spin-down subbands closer together, leading to a distinct valley-dependent rearrangement of the band structure compared with the WSe$_2$-based heterostructure.

Recent studies have shown that the proximity-induced exchange fields in monolayer MoSe$_2$ and WSe$_2$ arising from an adjacent ferromagnetic CrI$_3$ monolayer can be widely tuned by twisting and electrostatic gating~\cite{Zollner19,Zollner23}. In particular, the magnitude and even the sign of the exchange splittings depend sensitively on the twist angle between the TMDC and CrI$_3$ layers. Remarkably, the direction of the proximity-induced exchange field in the valence band undergoes a twist-angle-driven sign reversal. This reversal occurs near $8^\circ$ in WSe$_2$ and $16^\circ$ in MoSe$_2$, reflecting the strong geometric control of magnetic proximity effects in these heterostructures. This twist-angle induced sign reversal does not alter the spin character of the subbands: the exchange-induced splittings in both conduction and valence bands remain at least three orders of magnitude smaller than the intrinsic SOC, making the exchange field far too weak to flip the spin index. Consequently, in the WSe$_2$/CrI$_3$ heterostructure the two spin-subbands in either the conduction or valence band shift apart at the K valley and toward each other at the K' valley. In contrast, in the MoSe$_2$/CrI$_3$ heterostructure the conduction-band spin-subbands move closer together at the K valley and farther apart at the K' valley, while those of the valence band exhibit the opposite trend. Additionally, both the Fermi velocity $v_{\rm F}$ and the orbital gap $\Delta$ are modified by twisting; their magnitudes vary with the twist angle and may either increase or decrease accordingly. Moreover, the proximity-induced exchange parameters increase as the external electric field is tuned from negative to positive values, enabling effective gate control of the proximity exchange. We note, however, that neither the gap parameter nor the Fermi velocity is affected by the applied electric field.

Applying an in-plane electric field $\bm E = (E_x,E_y)$, parallel to the TMDC/FM interface, generates a transverse spin current in the TMDC layer due to its strong spin-orbit coupling. The interface-generated spin current originates from two complementary mechanisms, spin-orbit filtering and spin-orbit precession~\cite{Amin2018,manchon2019current}, and produces a net spin polarization at the interface. The spin polarization $\bm{S}={(2\pi)^{-2}}\sum_{\chi}\int{d^2\bm{k}} \bm{s}^{\chi}(\bm{k}) f(\varepsilon_{\bm{k}}^\chi)$ receives contributions from both intra-band and inter-band electronic processes. The intra-band component $\bm{S}^{oc}={(2\pi)^{-2}}\sum_{\chi}\int{d^2\bm{k}}\bm{s}^{\chi}(\bm{k})\delta f(\varepsilon_{\bm{k}}^{\chi})$ arises from changes in the electron occupation within the same band as carriers are accelerated by the electric field, with $\bm{s}^{\chi}(\bm{k}) =({\hbar}/{2})\langle\Psi_{\bm{k}}^\chi|\hat{\bm s}|\Psi_{\bm{k}}^\chi\rangle$ the spin expectation in the $\chi$-band with eigenvector $\Psi_{\bm{k}}^{\chi}$ and eigenvalue $\varepsilon_{\bm{k}}^{\chi}$, and $f(\varepsilon_{\bm{k}}^{\chi})$ the Fermi-Dirac distribution function. In contrast, the inter-band contribution $\bm{S}^{in}={(2\pi)^{-2}}\sum_{\chi}\int{d^2\bm{k}}\delta \bm{s}^{\chi}(\bm{k})f(\varepsilon_{\bm{k}}^{\chi})$ stems from field-induced modifications of the quasiparticle wave functions~\cite{Kurebayashi,Xiao17,Garate}, with $\delta\bm{s}^{\chi}(\bm{k}) =({\hbar}/{2})Re\langle\Psi_{\bm{k}}^\chi|\bm{\hat{\bm s}}|\delta\Psi_{\bm{k}}^\chi\rangle$. The total spin polarization, $\bm{S}=\bm{S}^{oc}+\bm{S}^{in}$, exerts a spin-orbit torque on the ferromagnetic layer, characterized by a local magnetization ${\bm{m}}$, as
\begin{equation}
\label{torque}
\bm{\tau}=\frac{2 J}{\hbar}{\bm{m}}\times\bm{S},
\end{equation}
where $J$ denotes the exchange energy. This torque is generally decomposed into two components according to their symmetry and antisymmetry with respect to the magnetization reversal, namely, the damping-like and the field-like terms. The extrinsic intra-band transitions primarily generate spin currents through scattering processes within a single band, and these currents typically contribute to field-like torques. In contrast, intrinsic inter-band processes involve mixing between different energy bands and are responsible for producing band structure-driven spin densities that give rise to damping-like torques. Importantly, the resulting spin-orbit torque can be sufficiently strong to switch the magnetization in TMDC/FM bilayers, underscoring its relevance for efficient spintronic control.

To evaluate the intra-band spin polarization, we first compute the spin expectation $\bm{s}^{\chi}(\bm{k}) =({\hbar}/{2})\langle\Psi_{\bm{k}}^\chi|\hat{\bm s}|\Psi_{\bm{k}}^\chi\rangle$ by using the eigenstates of the effective Hamiltonian in Eq. \eqref{H}. Denoting these eigenstates by
\begin{equation}
\label{psi}
|\Psi_{\bm{k}}^{\chi}\rangle=A_{\chi}\ e^{i \tau {\chi}_{_{\sigma}} k_x x} e^{i {\chi}_{\sigma} k_y y}
\left(
\begin{array}{c}
\tau {\chi}_{\sigma} C_{\chi}\ e^{- i\tau\theta}\\
1\\
\tau {\chi}_{\sigma} D_{\chi}\ e^{i\phi_m}\ e^{- i\tau\theta}\\
F_{\chi}\ e^{i\phi_m}
\end{array}
\right),
\end{equation}
the corresponding spin expectation components are obtained as
\begin{eqnarray}
s_x^{\chi}&=&{\hbar}\ |A_{\chi}|^2\ [\rm{Re}(F_{\chi}\ e^{i\phi_m})+ \rm{Re}(C_{\chi}^*D_{\chi}\ e^{i\phi_m})],\\
s_y^{\chi}&=&{\hbar}\ |A_{\chi}|^2\ [\rm{Im}(F_{\chi}\ e^{i\phi_m})+ \rm{Im}(C_{\chi}^*D_{\chi}\ e^{i\phi_m})],\\
s_z^{\chi}&=&{\hbar}\ |A_{\chi}|^2\ (|C_{\chi}|^2-|D_{\chi}|^2-|F_{\chi}|^2+1)/{2}.
\end{eqnarray}
Here, $F_{\chi}=F_1/F_2$ with $F_1=-[(a_c+\Delta/2-\varepsilon_{\bm{k}}^{\chi})(a_v-\Delta/2-\varepsilon_{\bm{k}}^{\chi})+a'_c a'_v-(\hbar v_{\rm F} |\bm{k}_{\chi}|)^2]$ and $F_2={a'_v(a_c+\Delta/2-\varepsilon_{\bm{k}}^{\chi})-a'_c(a_v+\Delta/2+\varepsilon_{\bm{k}}^{\chi})}$, $C_{\chi}=-({\hbar v_{\rm F} |\bm{k}_{\chi}|})^{-1}[{a_v-\Delta/2-\varepsilon_{\bm{k}}^{\chi}+a'_v F_{\chi}}]$, $D_{\chi}=-[{C_{\chi}(a_c+\Delta/2-\varepsilon_{\bm{k}}^{\chi})+\hbar v_{\rm F} |\bm{k}_{\chi}|}]/{a'_c}$, $a_{c(v)}=\tau\lambda_{c(v)}+B_{c(v)} \cos\theta_m$, $a'_{c(v)}=B_{c(v)} \sin\theta_m$, $\theta=\arctan(k_y/k_x)$, and  $A_{\chi}=({|C_{\chi}|}^2+{|D_{\chi}|}^2+{|F_{\chi}|}^2+1)^{-1/2}$. Note that the $\chi$-energy band is defined by $\chi=(\chi_s,\chi_{\sigma},\tau)$ in which $\chi_s=\pm 1$ denotes the two spin-subbands, and $\chi_{\sigma}=sgn(\varepsilon_{\bm{k}}^{\chi})=\pm 1$ refers to the conduction and valence band subspaces.

The n-th order non-equilibrium distribution function can be obtained within the relaxation time approximation, where the relaxation time is taken as momentum-independent, $\gamma(\bm{k})=\gamma$. In this framework, the distribution function follows from the recursive relations $f^{(n)}(\varepsilon_{\bm{k}}^{\chi})={e\gamma}{\hbar}^{-1}\bm{E}.\partial_{\bm{k}}{f^{(n-1)}(\varepsilon_{\bm{k}}^{\chi})}$ derived from the single-band steady-state Boltzmann equation
\begin{equation}
-\frac{e}{\hbar}\bm{E}.\nabla_{\bm{k}}f(\varepsilon_{\bm{k}}^{\chi})=-\frac{f(\varepsilon_{\bm{k}}^{\chi})-f^{(0)}(\varepsilon_{\bm{k}}^{\chi})}{\gamma(\bm{k})},
\end{equation}
allowing us to construct higher-order corrections systematically. Therefore, the intra-band spin polarization $\bm{S}^{oc}$ in response to an in-plane applied electric field can be obtained using
\begin{eqnarray}
\label{spin polarization}
\bm{S}^{oc}&=&(\frac{1}{2\pi})^2\sum_{\chi({\chi}_{\sigma}=+(-)1,\chi_s=\pm 1,\tau=\pm 1)}\int_0^\infty d\varepsilon\ |\bm{k}_{\chi}(\varepsilon)|\ \frac{\partial |\bm{k}_{\chi}(\varepsilon)|}{\partial \varepsilon}\nonumber\\
&\times&\bm{s}^{\chi}(\varepsilon)\int_{0}^{2\pi} d\theta\ \ [f^{(1)}(\theta,\varepsilon)+f^{(2)}(\theta,\varepsilon)],
\end{eqnarray}
with momentum-energy relation $|\bm{k}_{\chi}(\varepsilon)|=({2\hbar v_{\rm F}})^{-1} [4 a_c a_v + 4 a'_c a'_v + 4 {\varepsilon_{\bm{k}}^{\chi}}^2 - \Delta^2 +8 \chi_{s} (- {a'_c}^2 a_v^2 + 2 a_c a'_c a_v a'_v - a_c^2 {a'_v}^2 + {\varepsilon_{\bm{k}}^{\chi}}^2 [(a_c+a_v)^2 + ({a'_c}+{a'_v})^2]+ 4\varepsilon_{\bm{k}}^{\chi}\ \Delta\ (a_c^2 + {a'_c}^2- a_v^2 - {a'_v}^2) + \Delta^2 [(a_c-a_v)^2 + ({a'_c}-{a'_v})^2])^{1/2}]^{1/2}$, the non-equilibrium distribution functions $f^{(1)}(\theta,\varepsilon)=-\beta \gamma e|\bm E| (\mathrm{v}_x^{\chi}\cos\theta_E +\mathrm{v}_y^{\chi}\sin\theta_E) e^{\beta \varepsilon}[f^{(0)}(\varepsilon)]^2$ and $f^{(2)}(\theta,\varepsilon)=(\beta \gamma e|\bm E|)^2([(\partial \mathrm{v}_x^{\chi}/\partial k_x)\ {\cos^2\theta_E}+ (\partial \mathrm{v}_y^{\chi}/\partial k_y)\ \sin^2\theta_E+(\partial \mathrm{v}_x^{\chi}/\partial k_y+\partial \mathrm{v}_y^{\chi}/\partial k_x)\cos\theta_E \sin\theta_E](\partial f^{(0)}(\varepsilon)/\partial\varepsilon)+[\mathrm{v}_x^{\chi} \cos\theta_E +\mathrm{v}_y^{\chi}\sin\theta_E]^2(\partial^2 f^{(0)}(\varepsilon)/\partial\varepsilon^2))$, the excitation energy $\varepsilon=\varepsilon_{\bm{k}}^{\chi}-\mu$, the chemical potential $\mu$, $\bm{\mathrm{v}}^{\chi}=\chi_{\sigma}{[\hbar\ |\bm{k}_{\chi}(\varepsilon)|\partial |\bm{k}_{\chi}(\varepsilon)|/\partial \varepsilon]}^{-1}(k_x,k_y)=\chi_{\sigma}{[\hbar\ \partial |\bm{k}_{\chi}(\varepsilon)|/\partial \varepsilon]}^{-1}(\cos\theta,\sin\theta)$, $\theta_E=\arctan(E_y/E_x)$, $|\bm E|=\sqrt{E_x^2+E_y^2}$, and $\beta=(k_B T)^{-1}$ with $k_B$ the Boltzman constant, and $T$ the temperature.

The linear contribution $\bm{S}^{oc(1)}={(2\pi)^{-2}}\sum_{\chi}\int{d^2\bm{k}}\bm{s}^{\chi}(\bm{k}) f^{(1)}(\varepsilon_{\bm{k}}^{\chi})$ vanishes. For an isotropic dispersion, the angular dependence of the terms entering $f^{(1)}(\varepsilon_{\bm{k}}^{\chi})$ is proportional to $\sin\theta$ or $\cos\theta$, both of which yield zero upon angular integration. Consequently, the intra-band spin polarization $\bm{S}^{oc}$, and the associated spin-orbit torque, is the first leading term by the nonlinear contribution.

Then, we evaluate the contribution of the intrinsic inter-band transitions on the spin polarization in the linear response regime. Following the perturbation method to find the modifications in the wave function, we have~\cite{Kurebayashi,Xiao17,Garate}
\begin{equation}\label{S_in_2}
{\bm S}^{in}=\frac{e\hbar^2}{A} \sum_{\chi\neq\chi',\bf{k}}[f^{(0)}(\varepsilon^{\chi}_{\bf k})-f^{(0)}(\varepsilon^{\chi'}_{\bf k})]\frac{\rm{Im}\bigl[\langle \Psi^{\chi}_{\bf k}\vert{\hat{\bm s}}\vert\Psi^{\chi'}_{\bf k}\rangle\langle \Psi^{\chi'}_{\bf k}\vert{\bf{\rm{\hat{v}}\cdot E}}\vert\Psi^{\chi}_{\bf k}\rangle\bigr]}{(\varepsilon^{\chi}_{\bf k}-\varepsilon^{\chi'}_{\bf k})^2}
\end{equation}
for each valley  with $\rm{\hat{v}}=\hbar^{-1}\partial_{\bm{k}}\mathcal{H}$. We demonstrate that the contribution of the intrinsic inter-band transitions to the spin polarization is zero in the linear response regime [see Appendix \ref{appendix A}]. Furthermore, due to the much weaker magnitude of the intrinsic inter-band spin polarization and the associated spin-orbit torque in comparison with that produced by intra-band processes, the second-order response is negligible in this case and can be safely disregarded~\cite{zhou2022nonlinear,Majidi26}. Therefore, the linear spin polarization and the resulting SOT arising from the both of intra- and inter-band transitions will be zero and the nonlinear spin polarization stemming from the intra-band transitions exerts a nonlinear spin-orbit torque on the magnetization of the ferromagnetic layer.

\section{Numerical results and discussion}\label{results}
\begin{figure}[t]
\begin{center}
\includegraphics[width=3.5in]{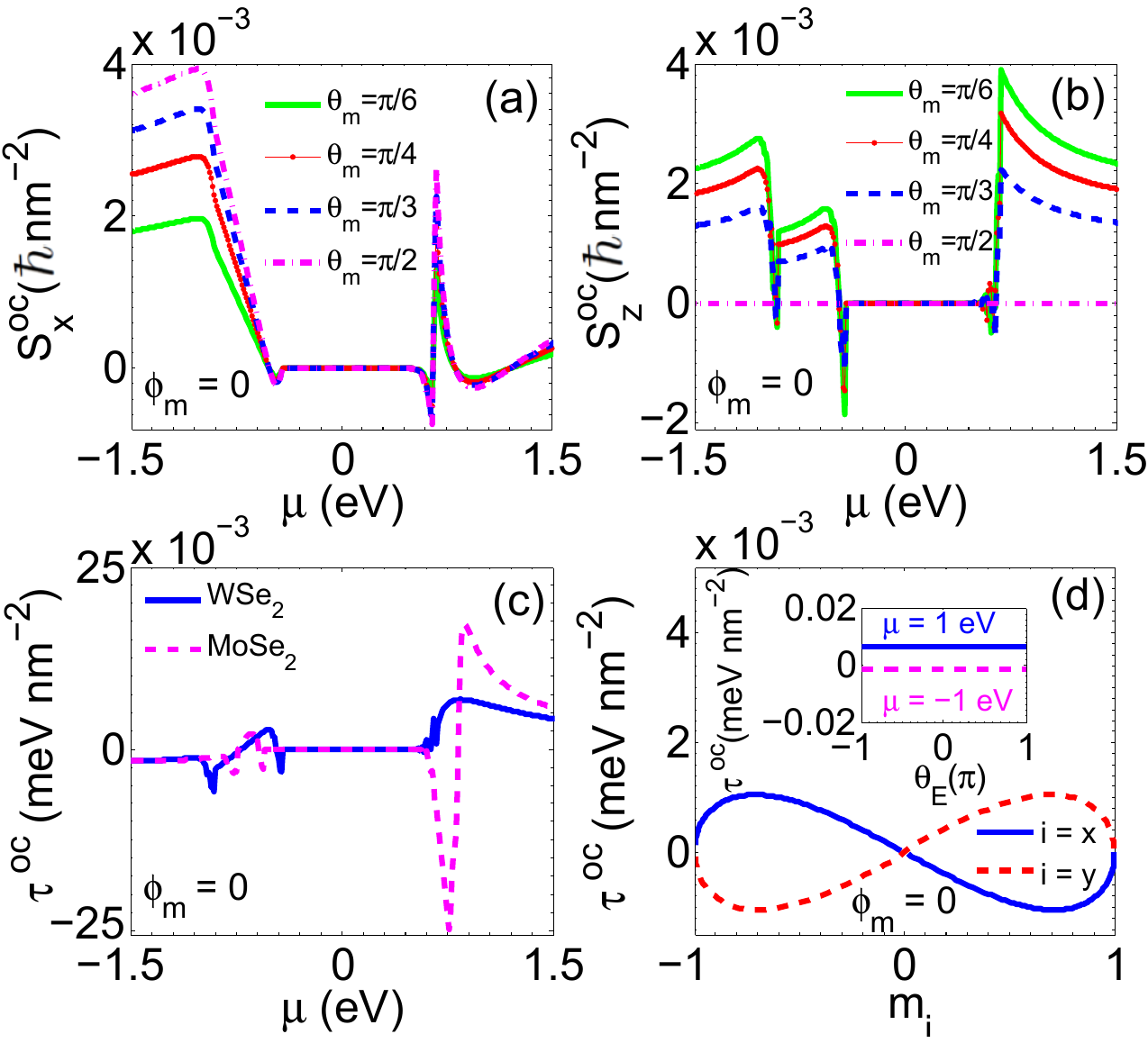}
\end{center}
\caption{\label{Fig:1} Chemical potential dependence of the non-zero components of the non-equilibrium spin polarization $\bm{S}^{oc}$ in WSe$_2$/CrI$_3$ bilayer with different magnetization directions $\theta_m$ [(a)-(b)] and the strength of the in-plane spin-orbit torque ${\tau}^{oc}$ in WSe$_2$- and MoSe$_2$-based bilayers with $\theta_M=\pi/4$ (c). (d) The strength of the spin-orbit torque ${\tau}^{oc}$ with respect to the non-zero components of the magnetization vector, when $\mu=-0.9$ eV. We have set the azimuthal angle $\phi_m=0$, the twist angle $\theta_{twist}=0^\circ$ and the in-plane electric field direction $\theta_E=\pi/4$. Inset of (d) presents the behavior of ${\tau}^{oc}$ in terms of $\theta_E$ for two values of the chemical potential, when $\theta_m=\pi/4$.}
\end{figure}

In this section, we present our numerical analysis of the nonlinear spin polarization and the resulting spin-orbit torque acting on the magnetization of the ferromagnetic layer. As discussed in Sec. \ref{Model}, the intrinsic inter-band transitions in addition to the intra-band transitions have no contribution to the spin polarization in linear response regime. We therefore compute the nonlinear spin polarization $\bm{S}^{oc}$ and the corresponding spin-orbit torque $\bm{\tau}^{oc}$, arising from the electron occupation changes within intra-band at room temperature, using Eqs. (\ref{spin polarization}) and (\ref{torque}), where the exchange parameter $J$ is replaced by $B_{c(v)}$ for the conduction (valence) band. Our analysis is performed in the limit of negligible Rashba spin-orbit coupling, as neither experimental data nor \textit{ab initio} studies presently offer a quantitative estimate for its magnitude in these heterostructures. This assumption further facilitates a direct comparison with the finite Rashba SOC results reported in Ref.~\onlinecite{Majidi26}.

Since including or excluding the SOC term in the bilayer calculations produces no significant change in the magnitude of the exchange parameters, we adopt the values reported in Ref.~\onlinecite{Zollner19} for the $30^{\circ}$-twisted WSe$_2$ (MoSe$_2$)/CrI$_3$ bilayer: $\Delta=1.417 (1.351)$ eV, $v_{\rm F}= 5.863\times10^5 (4.597\times10^5)$ m/s, $\lambda_{c}=13.81 (-9.675)$ meV, $\lambda_{v}=240.99 (94.43)$ meV, $B_c=-1.648 (-1.641)$ meV and $B_v=1.896 (0.502)$ meV. The magnitude of the proximity-induced exchange field $B_{v(c)}$ increases under twisting in WSe$_2$-based (MoSe$_2$-based) bilayers, whereas it decreases in the MoSe$_2$-based (WSe$_2$-based) bilayers. Importantly, twisting the CrI$_3$ layer relative to the TMDC reverses the sign of the valence-band exchange field, leading to an opposite sign of the valence-band spin-splitting in the absence of SOC. These trends highlight twisting as an effective means of tuning the proximity exchange field. We set the relaxation time to $\gamma = 3$ ps, the in-plane electric field to $E=0.2$ mV/nm, and $k_B T=25.7$ meV for room temperature. The chemical potential $\mu$ is expressed in electron volt (eV). For device-relevant conditions, single-layer and multilayer TMDCs may be $n$- or $p$-type doped to generate the desired carrier  densities~\cite{Radisavljevic,Fontana,Laskar}.

\begin{figure}[t]
\begin{center}
\includegraphics[width=3.4in]{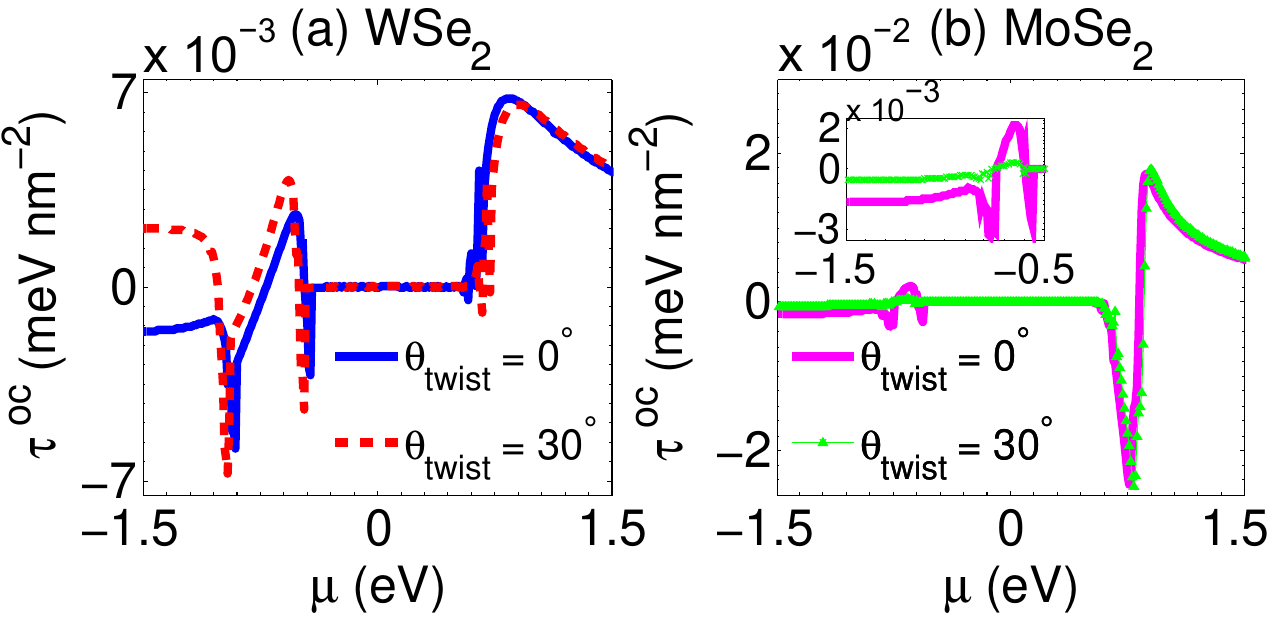}
\end{center}
\caption{\label{Fig:3} Left (Right) panel: The chemical potential dependence of the in-plane spin-orbit torque ${\tau}^{oc}$ for WSe$_2$- (MoSe$_2$-) based bilayer in the absence and presence of twisting (with $\theta_{twist}=30^\circ$), when $\theta_m=\pi/4$ and $\phi_m=0$. Inset of (b) shows the zoomed-in view of $\tau^{oc}$ in the case of p-doped bilayer.}
\end{figure}

Figures \ref{Fig:1}(a) and \ref{Fig:1}(b) present the non-zero components of the current-induced spin polarization $\bm{S}^{oc}$ as a function of the chemical potential $\mu$ for several magnetization orientations ($\theta_m,\phi_m=0$) in untwisted WSe$_2$/CrI$_3$ heterostructure, under an applied electric field direction $\theta_E=\pi/4$. The results reveal a strongly asymmetric dependence of both spin polarization components on $\mu$, characterized by sharp peaks, dips, and sign reversals within narrow energy windows for both n- and p-type doping. The magnitudes of $S_x$ and $S_z$ are comparable: $S_x$ dominates in the p-doped regime and increases with $\theta_m$, whereas $S_z$ is larger for n-type doping and vanishes as $\theta_m$ approaches $\pi/2$. Furthermore, the intrinsic band gap of the WSe$_2$/CrI$_3$ bilayer produces a corresponding gap in $\bm{S}^{oc}$ for $E_{\chi_{s}=1,\tau=-1}^v<\mu<E_{\chi_{s}=1,\tau=-1}^c$, with $E_{\chi_s,\tau}^{c(v)}=\pm (\Delta/2-\chi_s\sqrt{{\lambda^2}_{c(v)}+{B^2}_{c(v)}+2\tau\lambda_{c(v)}B_{c(v)}\cos\theta_m})$ the energy of the conduction (valence) band edge for the spin-$\chi_s$ subband of the $\tau$ valley.

\begin{figure}[]
\begin{center}
\includegraphics[width=3.6in]{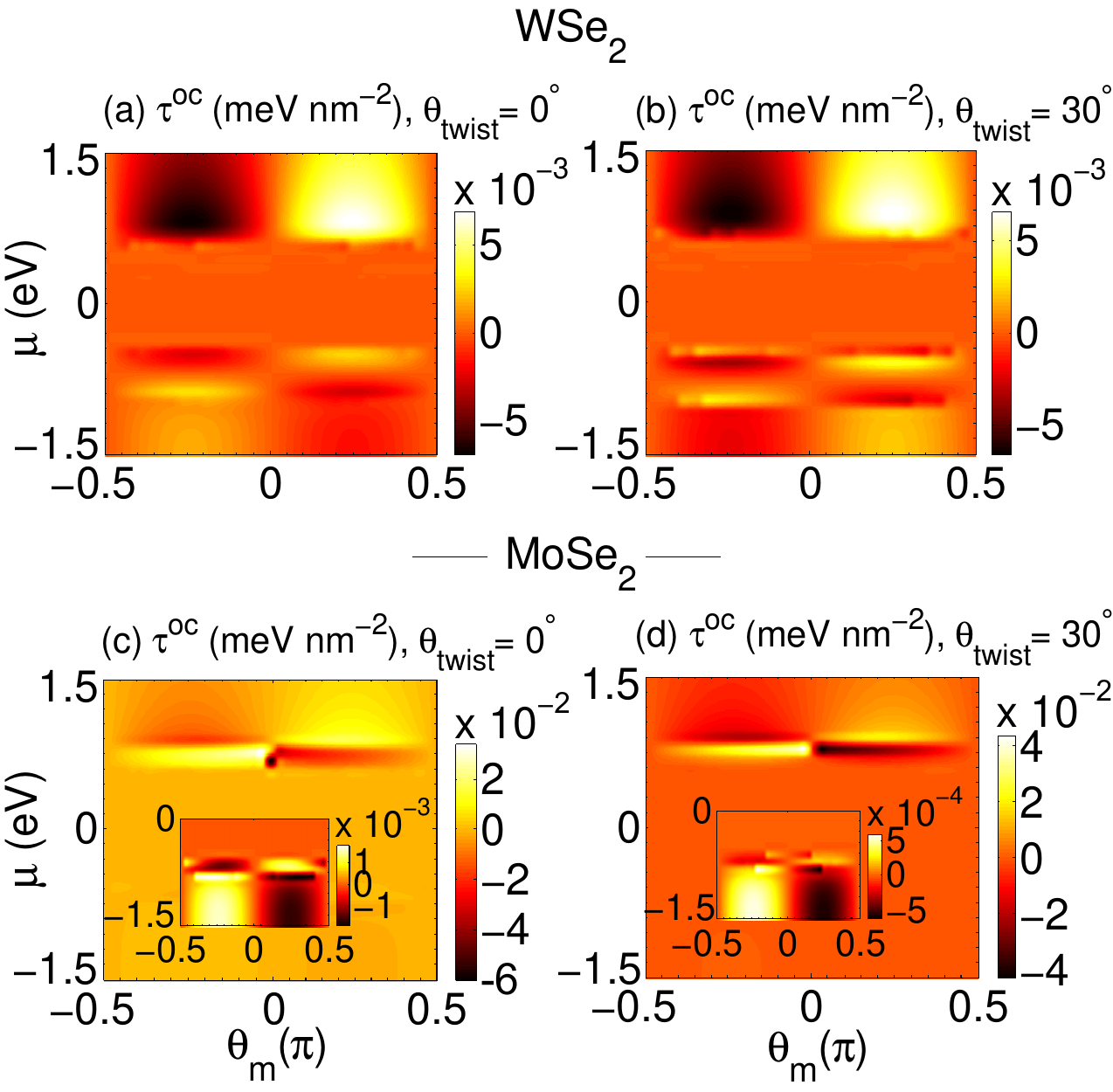}
\end{center}
\caption{\label{Fig:4} Top (Bottom) panel: The strength of the spin-orbit torque ${\tau}^{oc}$ in terms of the magnetization direction $\theta_m$ and the chemical potential $\mu$ in the absence and presence of twisting with $\theta_{twist}=30^\circ$ for WSe$_2$/CrI$_3$ [(a)-(b)] (MoSe$_2$/CrI$_3$ [(c)-(d)]) bilayer, when $\phi_m=0$.}
\end{figure}

The non-equilibrium spin polarization generated by the applied current exerts an in-plane spin-orbit torque $\bm{\tau}^{oc}=2 J \hbar^{-1}(S_z^{oc}\sin\theta_m-S_x^{oc}\cos\theta_m )\ \hat{y}$ on the magnetization of the ferromagnetic layer. As depicted in Fig. \ref{Fig:1}(c), the SOT also exhibits an asymmetric dependence on $\mu$, with larger amplitude in the n-doped regime and two sign reversals for p-type doping. Its magnitude decreases with increasing $\theta_m$ and vanishes at $\theta_m=\pi/2$ [see Fig. \ref{Fig:4}(a)]. Figure \ref{Fig:1}(d) displays the strength of the intra-band SOT as a function of the non-zero magnetization components for $\phi_m=0$. Importantly, the odd symmetry of $\bm{\tau}^{oc}$ with respect to $m_x$ and $m_z$ indicates that the intra-band SOT is of field-like character. We also show that the SOT is essentially insensitive to the direction of the in-plane electric field $\theta_E$ [see inset of Fig. \ref{Fig:1}(d)].

\begin{figure}[t]
\begin{center}
\includegraphics[width=3.6in]{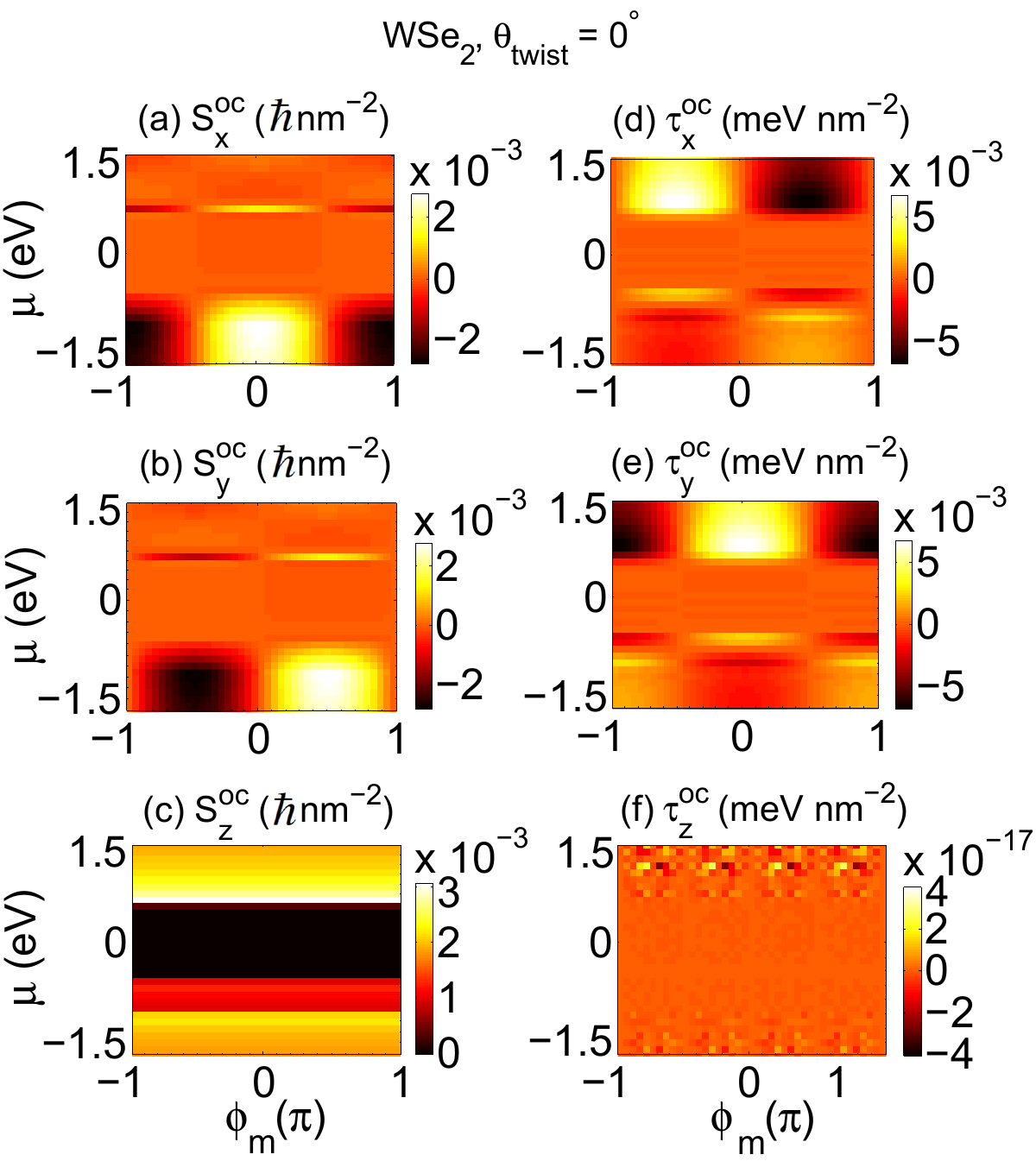}
\end{center}
\caption{\label{Fig:8} Left (Right) panel: The spin polarization (spin-orbit torque) components in terms of the magnetization direction $\phi_m$ and the chemical potential $\mu$ in untwisted WSe$_2$-based bilayer, when $\theta_m=\pi/4$.}
\end{figure}

We further examine how the choice of TMDC layer influences the generated SOT in Fig. \ref{Fig:1}(c). Replacing WSe$_2$ with MoSe$_2$ produces a pronounced enhancement of the SOT and introduces two sign reversals in the n-doped bilayer. Importantly, the chemical potential dependence of the intra-band spin polarization and the resulting SOT, particularly the width and amplitude of the associated peaks, is highly sensitive to both the sign and magnitude of the intrinsic spin-orbit coupling parameters [see Appendix \ref{appendix C}]. According to Ref.~\onlinecite{Zollner19}, the intrinsic SOC parameters in the conduction and valence bands of MoSe$_2$/CrI$_3$ (WSe$_2$/CrI$_3$) bilayers are $\lambda_c=-9.678$($13.81$) meV and $\lambda_v=94.43$($240.99$) meV. In the MoSe$_2$/CrI$_3$ bilayer, the negative $\lambda_c$ produces a sharp, high-amplitude peak in the spin polarization of each $\chi$-subband. This peak is primarily determined by the spin expectation value of carriers in the corresponding subband, and emerges at the energies where the spin-split subbands approach each other most closely in momentum space. The resulting opposite spin polarizations of the two spin-subbands then combine to produce a sign reversal of the spin polarization in the K and K' valleys, ultimately reversing the net spin polarization of the n-doped system. In contrast, the positive $\lambda_c$ in the WSe$_2$-based bilayer yields a broader, lower-amplitude peak in spin polarization, which suppresses the valley-resolved sign reversal and therefore prevents a reversal of the net spin polarization. Furthermore, the larger values of $\lambda_c$, and especially $\lambda_v$, in WSe$_2$ relative to MoSe$_2$ reduce the overall spin polarization, thereby enhancing the intra-band SOT in the MoSe$_2$-based bilayer~\cite{Majidi26}.

We next evaluate the effect of twisting on the generated SOT in Fig. \ref{Fig:3} for the magnetization angles $\theta_m=\pi/4$ and $\phi_m=0$. Twisting the TMDC layer relative to the CrI$_3$ layer with $\theta_{twist}= 30^{\circ}$ leads to either amplification or attenuation of the SOT, depending on the chemical potential, and introduces additional qualitative changes in the p-doped regime. In the WSe$_2$-based bilayer, a $30^{\circ}$ twist induces a pronounced sign reversal within the p-type doping regime. In contrast, the same twist suppresses the sign changing behavior across a broad range of chemical potentials in the p-doped MoSe$_2$-based bilayer, causing the SOT sign reversal to emerge only at smaller magnitudes of chemical potential in the p-doped region. The emergence or disappearance of these sign reversals in p-doped heterostructure is directly tied to the twist-angle induced sign change of the valence-band proximity exchange parameter $B_v$, which controls the spin-splitting and thus the valley-dependent spin polarization.

\begin{figure}[t]
\begin{center}
\includegraphics[width=3.6in]{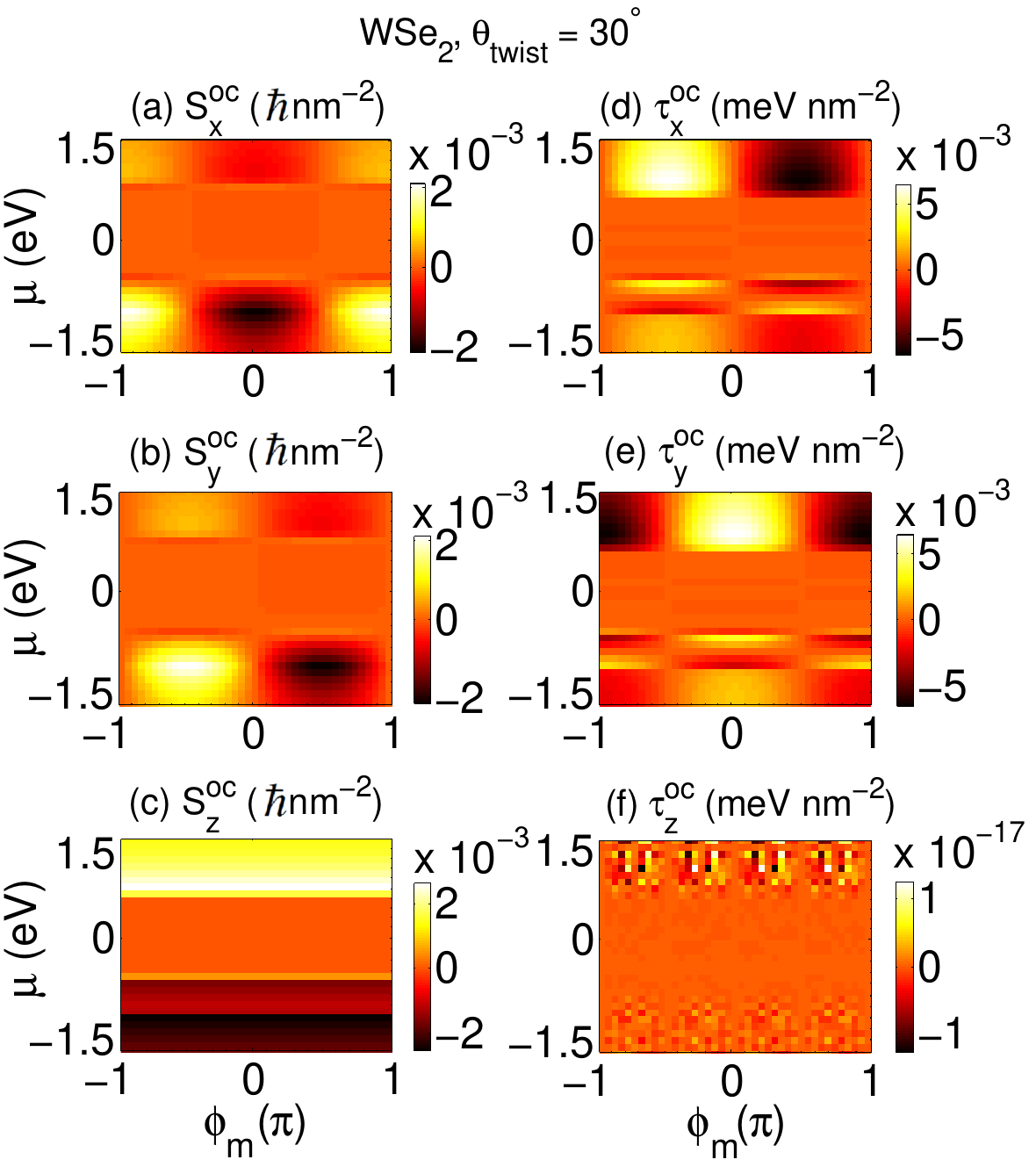}
\end{center}
\caption{\label{Fig:9} Left (Right) panel: The spin polarization (spin-orbit torque) components in terms of the magnetization direction $\phi_m$ and the chemical potential $\mu$ in twisted WSe$_2$-based bilayer with $\theta_{twist}=30^{\circ}$, when $\theta_m=\pi/4$.}
\end{figure}

To develop a qualitative understanding of how the SOT is influenced by the magnetization direction across a wide range of chemical potentials, we present colormaps of the strength of the non-zero SOT components as functions of $\theta_m$ ($\phi_m$) and the chemical potential $\mu$, for fixed $\phi_m$ ($\theta_m$), in the following three figures. Figure \ref{Fig:4} shows ${\tau}^{oc}(\theta_m,\mu)$ for WSe$_2$- and MoSe$_2$-based bilayers, in the absence and presence of twisting ($\theta_{twist}=30^\circ$), with the azimuthal angle fixed at $\phi_m=0$. The spin-orbit torque displays an odd dependence on $\theta_m$, attaining its maximum near $\theta_m=\pi/4$ and vanishing at $\theta_m=0,\pm \pi/2$. A pronounced asymmetry between n-type and p-type doping is evident in both TMDC/CrI$_3$ bilayers. In addition, the twist-induced sign reversal observed in p-doped WSe$_2$, contrasted with its absence for a broad range of chemical potential in p-doped MoSe$_2$, highlights the material-dependent nature of the torque. We should note that the twist-angle driven modifications of the proximity-exchange parameters and the gap parameter $\Delta$ strongly influence the $\chi$-band edge energies, thereby shaping the zero-SOT gap and the positions of the SOT peaks. Most importantly, the strength of the SOT in the MoSe$_2$-based bilayer exceeds that of the WSe$_2$-based bilayer by an order of magnitude. Even more striking is that the nonlinear intra-band contribution itself is an order of magnitude larger than the linear SOT reported in the presence of Rashba SOC~\cite{Majidi26}.

Figures \ref{Fig:8} and \ref{Fig:9} show the components of the current-induced spin polarization (left panels) and the resulting SOT (right panels) as functions of $\phi_m$ and $\mu$ for the WSe$_2$-based bilayer, respectively without and with twisting ($\theta_{twist}=30^{\circ}$), at a fixed polar angle $\theta_M =\pm\pi/4$. The in-plane spin polarization components, $S_x$ and $S_y$, exhibit a strongly asymmetric dependence on the chemical potential, with significantly larger magnitudes in the p-doped regime, in contrast to the out of plane component $S_z$. This behavior produces an in-plane SOT with a large discrepancy between n-type and p-type doping: the SOT components are large in the n-doped regime and undergo sign reversal in the p-doped regime. Tuning the azimuthal angle $\phi_m$ at fixed $\theta_m$, the $x$ and $y$ components of SOT exhibit sign reversals at $\phi_m=0,\pm\pi$ and $\phi_m=\pm\pi/2$, respectively. Comparing these results with those of the twisted structure shows that twisting leads to attenuation of the spin polarization magnitude and amplification of the SOT in the p-doped regime, accompanied by sign changes in their components [see Fig. \ref{Fig:9}]. We should note that reducing the temperature below room temperature does not qualitatively alter the chemical potential dependence of the generated SOT components, but merely reduces their magnitudes. We further find that lowering the temperature below room temperature does not modify the qualitative dependence of the generated SOT components on the chemical potential; it only diminishes their magnitudes.

\begin{figure}[t]
\begin{center}
\includegraphics[width=3.4in]{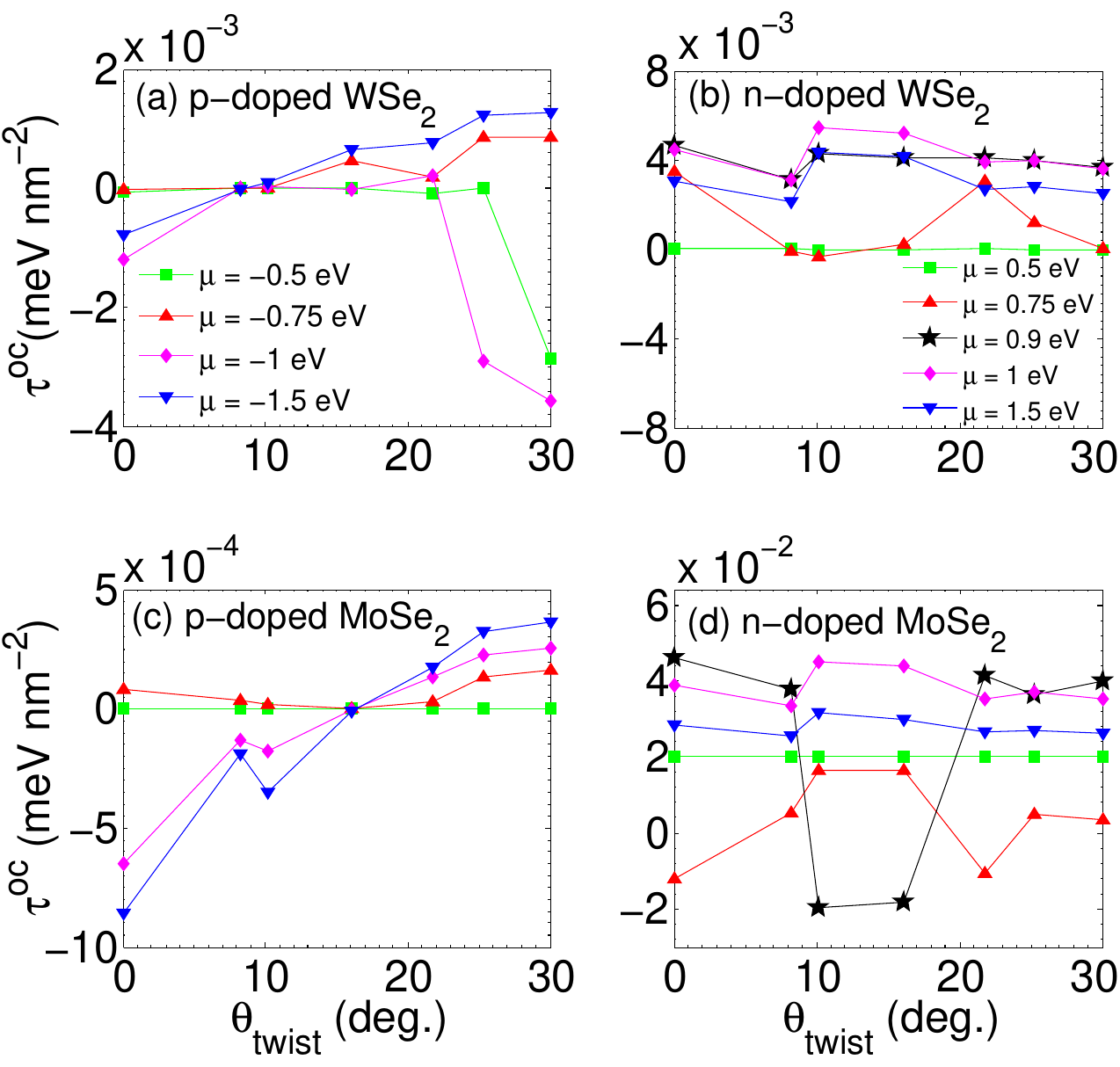}
\end{center}
\caption{\label{Fig:5} Top (Bottom) panels: The twist-angle dependence of the strength of the SOT for the WSe$_2$ (MoSe$_2$)/CrI$_3$ bilayer with n- and p-type doping, when $\theta_m =\pi/4$ and $\phi_m=0$. The labels in panels (c) and (d) are the same as those introduced in panels (a) and (b), respectively.}
\end{figure}

\begin{figure}[t]
\begin{center}
\includegraphics[width=3.4in]{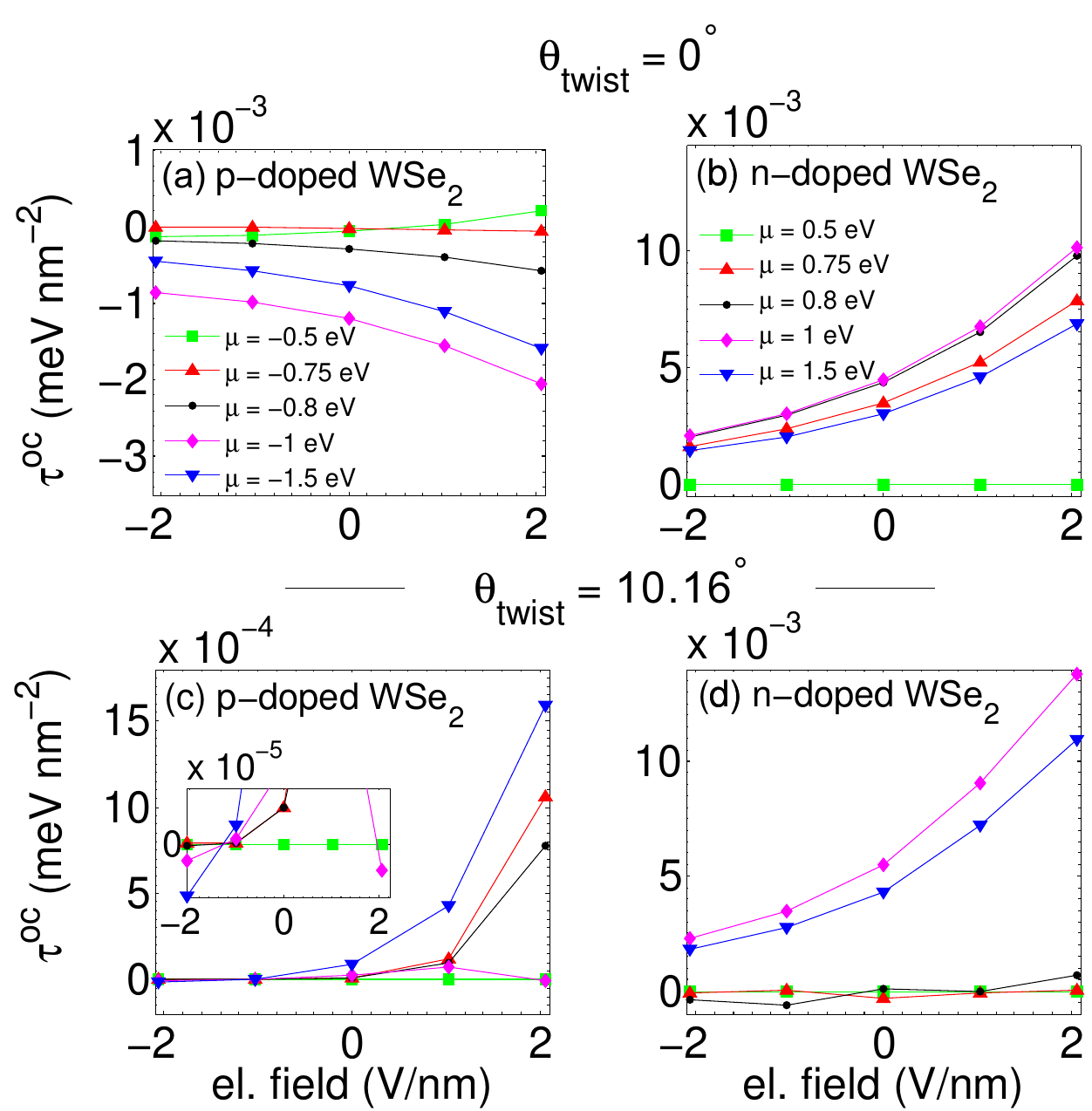}
\end{center}
\caption{\label{Fig:6} The behavior of the strength of the SOT as a function of the gate electric field for different values of the chemical potential in WSe$_2$/CrI$_3$ bilayer with $\theta_{twist}=0^\circ$ (a-b) and $10.16^\circ$ (c-d), when $\theta_m =\pi/4$ and $\phi_m=0$. The left (right) panels are for p-(n-) type doping. The labels in panels (c) and (d) are the same as those introduced in panels (a) and (b), respectively.}
\end{figure}
\begin{figure}[t]
\begin{center}
\includegraphics[width=3.6in]{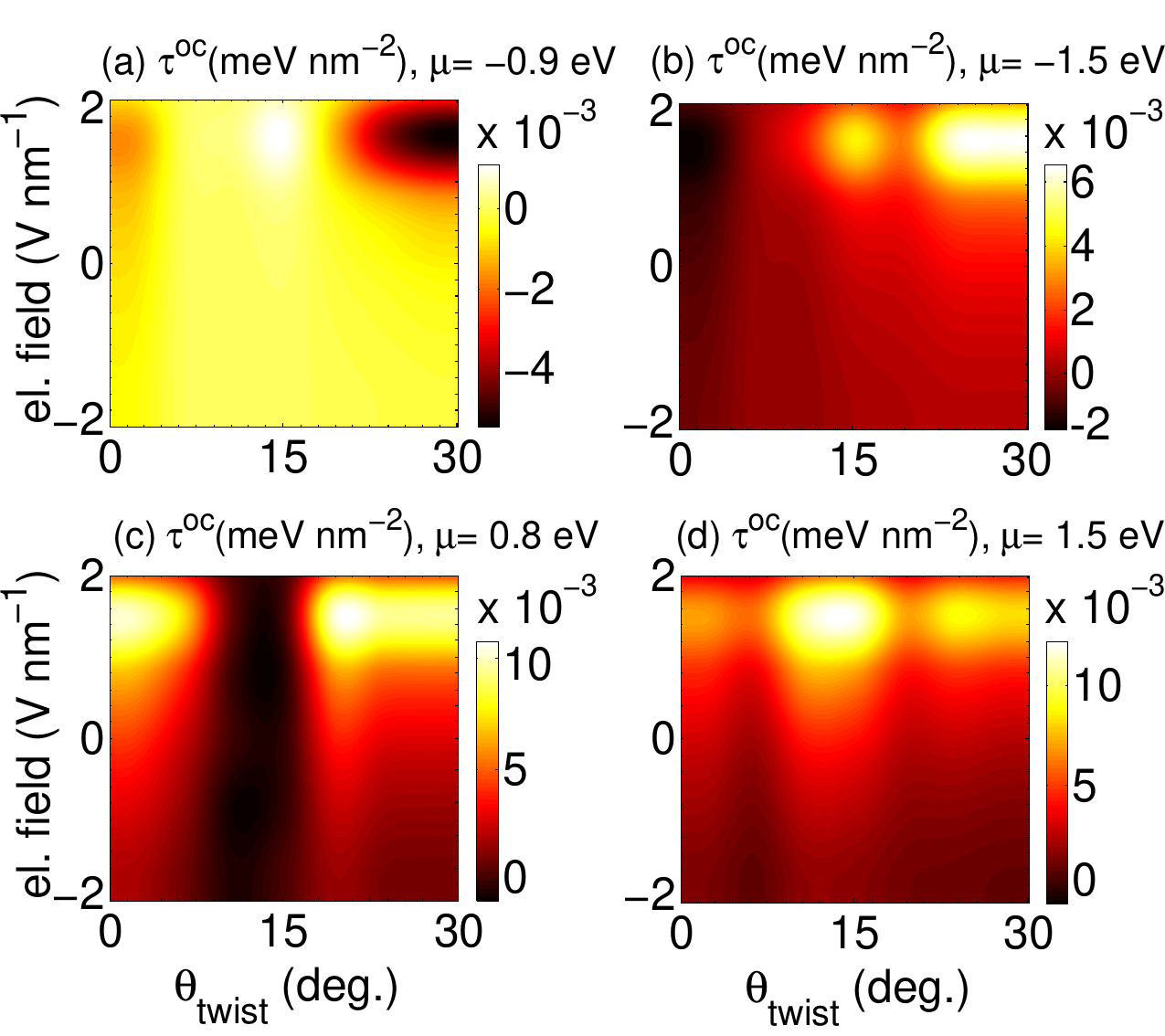}
\end{center}
\caption{\label{Fig:11} Top (Bottom) panel: The strength of the SOT as functions of the twist angle and the gate electric field for different values of the chemical potential in WSe$_2$/CrI$_3$ bilayer with p-type (a-b) and n-type (c-d) doping, when $\theta_m =\pi/4$ and $\phi_m=0$. The plots are interpolated from the data obtained in Ref. \onlinecite{Zollner23}.}
\end{figure}

A recent detailed analysis of twist-angle-dependent proximity exchange at intermediate angles shows that the band-edge splittings can vary substantially with the relative twist angle, and that the spin ordering may even reverse in the absence of SOC~\cite{Zollner23}. For twists between $0^\circ$ to $30^\circ$, the TMDC conduction-band edge splitting remains nearly constant at approximately $-3$ meV. In contrast, the TMDC valence-band edge exhibits an almost linear evolution of the spin splitting with twist angle, ranging from $-2$ to $2$ meV. The twist angle at which the spin ordering switches is found to be around $8^\circ$ for WSe$_2$ and $16^\circ$ for MoSe$_2$. Moreover, applying a transverse electric field of only a few V/nm across the twisted TMDC/CrI$_3$ bilayers enables a giant tunability of the proximity-induced exchange coupling.

Motivated by this, we highlight the extraordinary tunability of the SOT by the effects of the twist-angle and the gate electric-field in the following figures. We use the parameters obtained in Ref. \onlinecite{Zollner23} for seven different twist angles between $0^\circ$ and $30^\circ$, and interpolate the data of Fig. \ref{Fig:11} using "bicubic" interpolation method. As displayed in Fig. \ref{Fig:5}, twisting the TMDC layer relative to the ferromagnetic CrI$_3$ strongly modulates the SOT in WSe$_2$- and MoSe$_2$-based bilayers. This behavior arises from the twist-driven evolution of the proximity-induced exchange parameters $B_c$ and $B_v$, which directly modulate the SOT amplitude and sign [see Eq. \eqref{torque}], the zero-SOT gap and the $\chi$-band edge energies, all of which track the twist-angle dependence. Both p-doped WSe$_2$/CrI$_3$ and p-doped MoSe$_2$/CrI$_3$ bilayers exhibit SOT sign reversals, coinciding with the twist-induced sign change of the valence-band exchange parameter $B_v$. In p-doped WSe$_2$/CrI$_3$ bilayers, lowering the chemical potential substantially enhances the SOT magnitude. For n-type doping, varying the twist angle between $0^\circ$ and $30^\circ$ also produces a clear SOT sign reversal in both TMDCs, depending on the chemical potential. A pronounced discrepancy between the two TMDCs is evident, with the torque magnitude in n-type doped MoSe$_2$ exceeding that of WSe$_2$ by more than an order of magnitude. This difference primarily originates from the opposite signs and different magnitudes of their conduction-band spin-orbit coupling parameters. Moreover, tuning the chemical potential at a fixed twist angle can introduce or eliminate the sign reversal, with the outcome set by the TMDC material and doping regime.

Most importantly, tuning the gate electric field from $-2$ to $2$ V/nm strongly modifies the SOT, producing both substantial amplification and sign reversals depending on the chemical potential [see Fig. \ref{Fig:6}]. This response originates from the direct influence of the gate electric field on the proximity-induced exchange parameters $B_c$ and $B_v$, whose magnitudes increase under gating~\cite{Zollner23}. In the absence of twisting, gating alone can reverse the SOT sign in p-doped WSe$_2$ bilayers. When twisting is present, additional sign reversals appear for both p-type and n-type doping. For the specific twist angle $\theta_{twist} = 10.16^{\circ}$, the gate field even drives a sign reversal in $B_v$, and the SOT tracks these changes accordingly. Notably, the torque magnitude in n-type doped bilayers can exceed that of p-type systems by more than an order of magnitude.

The combined effects of the twist-angle and the gate electric field on the SOT can be apparently seen from Fig. \ref{Fig:11} for WSe$_2$/CrI$_3$ bilayer with p-type (a-b) and n-type (c-d) doping, when the magnetization direction is set to $\theta_m =\pi/4$ and $\phi_m=0$. The overall response is strongly controlled by the chemical potential. For p-doped bilayer, the SOT exhibits sign reversals over a wide twist angle window, and the application of the gate electric field enhances the torque magnitude while shifting these reversal regions. In contrast, for n-doped case, the SOT remains predominantly positive and increases monotonically with field strength, except for a very narrow twist-angle interval where a sign reversal occurs. Importantly, the SOT amplitude in the n-type regime is roughly an order of magnitude larger than in the p-type case, with the strongest response appearing at $\mu=1.5$ eV. These results demonstrate that both twist angle and chemical potential serve as effective control parameters for tuning the magnitude and sign of SOT in the system.

Comparing the results with those obtained in finite-Rashba regime~\cite{Majidi26} reveals a qualitative shift in the spin-orbit physics of TMDC/CrI$_3$ heterostructures. With finite Rashba SOC, intrinsic inter-band and extrinsic intra-band mechanisms both contribute in the linear response regime, producing field-like and damping-like torques with $\tau_{FL}/\tau_{DL}$ ratios of $10^3$ in WSe$_2$ and $10$ in MoSe$_2$. Sign reversals occur only for the field-like torque in n-doped MoSe$_2$ and for the damping-like torque in n-doped WSe$_2$ and MoSe$_2$, with tuning the chemical potential. MoSe$_2$/CrI$_3$ bilayer exhibits a three order of magnitude enhancement of the damping-like torque and an order of magnitude increase of the field-like torque relative to WSe$_2$-based bilayer. Increasing Rashba SOC strengthens both torque components in WSe$_2$ and p-type MoSe$_2$, while suppressing the damping-like term and eliminating the field-like sign reversal in n-doped MoSe$_2$. In sharp contrast, when Rashba SOC is set to zero, all linear response contributions vanish and the nonlinear intra-band contribution produces a purely field-like torque that is an order of magnitude larger than in the finite-Rashba case, with sign reversals in both n-type and p-type systems. MoSe$_2$/CrI$_3$ bilayer shows an additional order of magnitude enhancement and two sign reversals in the n-doped case.

In the finite-Rashba regime, twisting the TMDC layer can induce sign reversals in MoSe$_2$- and p-doped WSe$_2$-based heterostructures, while a transverse gate electric field provides strong tunability in n-doped WSe$_2$-enhancing the torque by nearly an order of magnitude- and producing a sign flip in p-doped structure at a twist angle of $10.16^{\circ}$. In the zero-Rashba limit, twisting can introduce or suppress sign reversals across both n-type and p-type WSe$_2$ and MoSe$_2$, and electrostatic gating enables nearly an order of magnitude modulation of the torque in n-doped WSe$_2$/CrI$_3$ while flipping its sign in p-type systems and in twisted n-type and p-type structures.

\section{conclusion}\label{conclusion}

In summary, we have developed a comprehensive microscopic theory of current-induced spin-orbit torque (SOT) in transition metal dichalcogenide (TMDC) and chromium iodide (CrI$_3$) van der Waals bilayers, with WSe$_2$ and MoSe$_2$ as TMDC, revealing that nonlinear intra-band mechanism is responsible for generating spin polarization and SOT in these systems in the limit of zero Rashba spin-orbit coupling (SOC). By explicitly demonstrating that both intrinsic inter-band and extrinsic intra-band mechanisms yield vanishing spin polarization in the linear response regime, a suppression enforced by symmetry, our work establishes a fundamental departure from conventional SOT mechanisms in heavy metal, topological insulator, and Rashba interfaces. This result highlights the unique role of nonlinear carrier dynamics in TMDC-based heterostructures and underscores the importance of going beyond linear response to correctly capture their spin-orbit physics.

The resulting nonlinear intra-band spin polarization generates a purely in-plane field-like torque whose magnitude, sign, and symmetry depend sensitively on the chemical potential, doping type, and magnetization orientation. We have found pronounced asymmetries between n-type and p-type doping, along with multiple sign reversals upon tuning the chemical potential, and a stronger torque in the n-type case. A key outcome of our analysis is the strong material dependence of the torque: an n-doped MoSe$_2$/CrI$_3$ bilayer exhibits a nonlinear SOT up to an order of magnitude larger than that of WSe$_2$/CrI$_3$, together with two sign changes. This behavior originates from the negative sign and smaller magnitude of the conduction-band SOC parameter in MoSe$_2$. These findings demonstrate that TMDC selection provides a powerful design knob for optimizing torque efficiency in ultrathin spintronic devices.

We further show that structural and electrostatic control, specifically twisting and gating, provide highly effective external tuning parameters. Twisting the TMDC layer relative to the monolayer CrI$_3$, by angles between $0^{\circ}$ and $30^{\circ}$, modifies the proximity-induced exchange fields and can either introduce or suppress sign reversals in the torque of both WSe$_2$- and MoSe$_2$-based heterostructures, depending on the chemical potential. Importantly, twisting can also lead to a significant amplification of the SOT in p-type doped WSe$_2$-based bilayer. Electrostatic gating provides an equally powerful and versatile control mechanism: by increasing the proximity exchange fields, a transverse gate electric field sweep from $-2$ to $2$ V/nm enables nearly an order of magnitude modulation of the SOT of n-doped WSe$_2$/CrI$_3$ bilayer and can even flip its sign in p-type doped structures. Crucially, this remarkable tunability, both in magnitude and sign, not only persists in p-type structures but also extends to n-type systems across a broad range of twist angles under applied gate fields, underscoring the robustness and flexibility of these control pathways.

Together, our results position TMDC/CrI$_3$ heterostructures as systems with an unusually rich and highly controllable SOT landscape, offering multiple independent tuning parameters: doping, twist angle, gate electric field for device engineering, and establish nonlinear intra-band transport as the central mechanism governing next-generation low-power SOT devices. Collectively, these findings show that the zero-Rashba limit reveals a fundamentally different SOT mechanism, dominated entirely by nonlinear intra-band transitions, in sharp contrast to the mixed inter-band and intra-band response that governs the finite-Rashba regime.

\appendix

\section {Intrinsic spin-orbit torque from inter-band transitions}\label{appendix A}

We calculate the contribution of the intrinsic inter-band transitions on the spin polarization in the linear response regime, by making use of Eq. (\ref{S_in_2}). Applying $\sum_{\bm k}={(2\pi)^{-2}}{A}\int |\bm{k}| d|\bm{k}|\ d\theta$, the $x$-component of the spin polarization ${\bf S}^{in}_{+++}$ for the $\chi_s=+1$ spin-subband of electrons in the conduction-band (with ${\chi}_{\sigma}=+1$) of K valley (with $\tau=+1$) will be obtained as
\begin{widetext}
\begin{eqnarray}
{S}^{in,x}_{+++}&=&\frac{e\hbar^2}{4\pi^2} \int_0^\infty d\varepsilon\ |\bm{k}_{\chi}(\varepsilon)| \frac{\partial |\bm{k}_{\chi}(\varepsilon)|}{\partial \varepsilon}\int_{0}^{2\pi} d\theta\ \lbrace [f(\varepsilon^{+++}_{\bf k})-f(\varepsilon^{+-+}_{\bf k})]\frac{\rm{Im}\bigl[\langle \Psi^{+++}_{\bf k}\vert{\hat{s}_x}\vert\Psi^{+-+}_{\bf k}\rangle\langle \Psi^{+-+}_{\bf k}\vert{\bf{\rm{\hat{v}}\cdot E}}\vert\Psi^{+++}_{\bf k}\rangle\bigr]}{(\varepsilon^{+++}_{\bf k}-\varepsilon^{+-+}_{\bf k})^2}\nonumber\\
&+&[f(\varepsilon^{+++}_{\bf k})-f(\varepsilon^{-++}_{\bf k})]\frac{\rm{Im}\bigl[\langle \Psi^{+++}_{\bf k}\vert{\hat{s}_x}\vert\Psi^{-++}_{\bf k}\rangle\langle \Psi^{-++}_{\bf k}\vert{\bf{\rm{\hat{v}}\cdot E}}\vert\Psi^{+++}_{\bf k}\rangle\bigr]}{(\varepsilon^{+++}_{{\bf k}}-\varepsilon^{-++}_{{\bf k}})^2}+[f(\varepsilon^{+++}_{{\bf k}})-f(\varepsilon^{--+}_{{\bf k}})]\frac{\rm{Im}\bigl[\langle \Psi^{+++}_{{\bf k}}\vert{\hat{s}_x}\vert\Psi^{--+}_{{\bf k}}\rangle\langle \Psi^{--+}_{{\bf k}}\vert{\bf{\rm{\hat{v}}\cdot E}}\vert\Psi^{+++}_{{\bf k}}\rangle\bigr]}{(\varepsilon^{+++}_{{\bf k}}-\varepsilon^{--+}_{{\bf k}})^2}\rbrace.\nonumber\\
\end{eqnarray}
The imaginary part of the first term in rhs of the above equation will be
\begin{equation}
\begin{split}
&\rm{Im}\bigl[\langle \Psi^{+++}_{{\bf k}}\vert{\hat{s}_x}\vert\Psi^{+-+}_{{\bf k}}\rangle\langle \Psi^{+-+}_{{\bf k}}\vert{\bf{\rm{\hat{v}}\cdot E}}\vert\Psi^{+++}_{{\bf k}}\rangle\bigr]=\\
&v_{\rm{F}}(A_{+++}A_{+-+})^2[N_{+++}P_{+-+}+N_{+-+}P_{+++}+M_{+++}+M_{+-+}][M_{+++}P_{+-+}-M_{+-+}P_{+++}-N_{+++}+N_{+-+}](E_x\sin\theta-E_y\cos\theta).
\end{split}
\end{equation}
For the second term we will have
 \begin{equation}
\rm{Im}\bigl[\langle \Psi^{+++}_{{\bf k}}\vert{\hat{s}_x}\vert\Psi^{-++}_{{\bf k}}\rangle\langle \Psi^{-++}_{{\bf k}}\vert{\bf{\rm{\hat{v}}\cdot E}}\vert\Psi^{+++}_{{\bf k}}\rangle\bigr]=v_{\rm{F}}(A_{+++})^4[-2N_{+++}P_{+++}+2M_{+++}][-2M_{+++}P_{+++}-2N_{+++}](E_x\sin\theta-E_y\cos\theta).
\end{equation}
Note that $(A,P,M,N)_{-++}=(A,P,M,N)_{+++}$. Finally the third term reads as
\begin{equation}
  \begin{split}
&\rm{Im}\bigl[\langle \Psi^{+++}_{{\bf k},c}\vert{\hat{s}_x}\vert\Psi^{--+}_{{\bf k},v}\rangle\langle \Psi^{--+}_{{\bf k},v}\vert{\bf{\rm{\hat{v}}\cdot E}}\vert\Psi^{+++}_{{\bf k},c}\rangle\bigr]=\\
&v_{\rm{F}}(A_{+++}A_{--+})^2[-N_{+++}P_{--+}-N_{--+}P_{+++}+M_{+++}+M_{--+}][M_{+++}P_{--+}+M_{--+}P_{+++}+N_{+++}+N_{--+}](-E_x \sin\theta+E_y\cos\theta).\\
\end{split}
\end{equation}
\end{widetext}

Since the imaginary terms appearing in ${\bm S}^{in}$ are proportional to $\sin\theta$ or $\cos\theta$ and the dispersion relation is isotropic in this system, the integration over $\theta$ is equal to zero. All other terms are qualitatively the same. Therefore, the contribution of the intrinsic inter-band transitions to the spin polarization and the corresponding torque $\bm{\tau}^{in}={2 J}({\hbar})^{-1}({\bm{m}}\times\bm{S}^{in})$ will be zero.

\section {Nonlinear extrinsic intra-band spin polarization: contribution of the carriers from distinct subbands}\label{appendix C}
\begin{figure}[t]
\begin{center}
\includegraphics[width=3.5in]{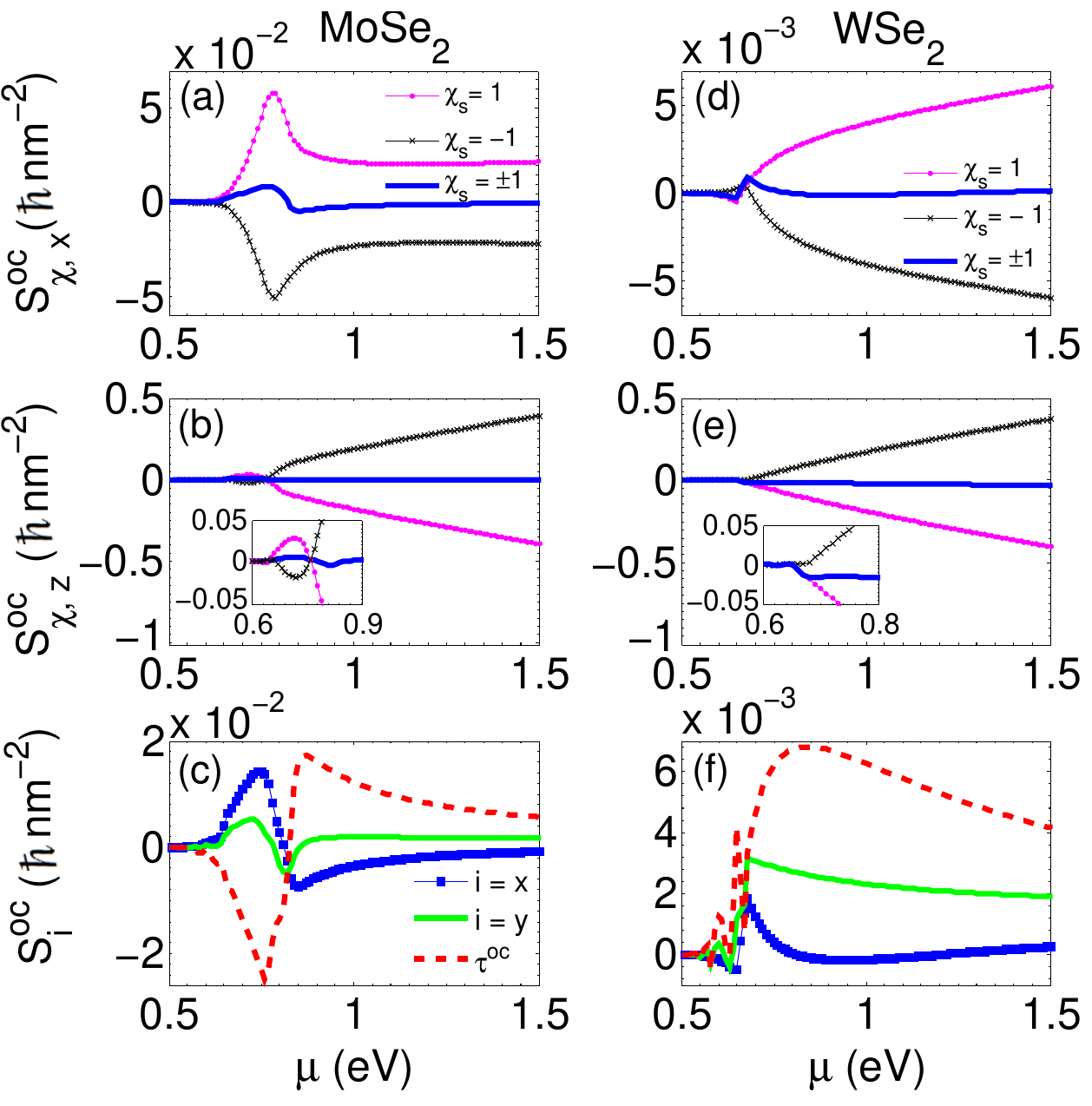}
\end{center}
\caption{\label{Fig:12} (Color online) Left (Right) panel: Chemical potential dependence of the spin polarization components ${S}_{\chi,x}^{oc}$ and ${S}_{\chi,z}^{oc}$ of the two $\chi_s=\pm 1$ subbands within the K valley (with $\chi_{\tau}= 1$) of the MoSe$_2$/CrI$_3$ [(a)-(b)] and WSe$_2$/CrI$_3$ [(d)-(e)] bilayers, when $\theta_m=\pi/4$, $\theta_{twist}=0^\circ$ and $\theta_E=\pi/4$. The total spin polarization components $S_x^{oc}$ and $S_z^{oc}$, and the generated spin-orbit torque in MoSe$_2$- and WSe$_2$-based bilayers are respectively presented in (c) and (f). Insets of (b) and (e) present the zoomed-in view of the $\chi$-subband spin polarization component $S_{\chi,z}^{oc}$. The labels in panels (b) and (e) are the same as those introduced in panels (a) and (d), respectively.}
\end{figure}

To clarify how the intra-band spin-orbit torque $\bm{\tau}^{oc}$ evolves with the chemical potential in TMDC/CrI$_3$ bilayers, we analyze the contribution of carriers from individual subbands to the spin polarization vector $\bm{S}^{oc}$. For a local magnetization ${\bm{m}}=(\sin\theta_m\cos\phi_m,\sin\theta_m\sin\phi_m,\cos\theta_m)$, lying in the x-z plane for $\phi_m=0$, the intra-band torque reduces to
$\bm{\tau}^{oc}=2 J \hbar^{-1}(S_z^{oc}\sin\theta_m-S_x^{oc}\cos\theta_m )\ \hat{y}$, showing that the generated torque is governed by the behavior of both $S_x^{oc}$ and $S_z^{oc}$. The left and right panels of Fig.~\ref{Fig:12} illustrate how the $\chi_s=\pm 1$ subbands in the K and K' valleys ($\chi_{\tau}=\pm 1$) contribute to the spin polarization components in n-doped MoSe$_2$/CrI$_3$ and WSe$_2$/CrI$_3$ bilayers, respectively, when ${\theta}_m=\pi/4$ and ${\theta}_{twist}=0^{\circ}$. In both systems, the two spin-split subbands within each valley produce spin polarizations of opposite sign as well as a pronounced peak in n-doped regime, as shown in Figs.~\ref{Fig:12}(a)-\ref{Fig:12}(b) and \ref{Fig:12}(d)-\ref{Fig:12}(e). This peak is primarily determined by the spin expectation value of carriers in the corresponding subband. Notably, the structure of the peak is highly sensitive to both the sign and magnitude of the intrinsic spin-orbit coupling parameters.

In MoSe$_2$/CrI$_3$, the negative conduction-band SOC parameter $\lambda_c$ produces a sharp, high amplitude peak-particularly in the x-component of the spin polarization for each $\chi$-subband. The strong opposition between the $\chi_s=\pm 1$ contributions results in a sign reversal of the valley-resolved spin polarization and, consequently, of the net spin polarization components and the associated torque in the n-doped regime [see Fig.~\ref{Fig:12}(c)]. In contrast, the positive and larger $\lambda_c$ in WSe$_2$/CrI$_3$ produces a broader and significantly weaker peak in the spin polarization components of the $\chi_s$-subbands. This broadening suppresses the valley-resolved sign reversal and the resulting torque [see Fig.~\ref{Fig:12}(f)], yielding a qualitatively different chemical potential dependence compared to MoSe$_2$-based bilayers.

%
\end{document}